\documentclass{article}

\usepackage{PRIMEarxiv}
\usepackage{amsmath} % for \overset
\DeclareMathOperator{\sign}{sign}
\usepackage{mathtools} % for \argmin
\DeclareMathOperator*{\argmin}{arg\,min}
\DeclareMathOperator*{\argmax}{arg\,max}
\usepackage{amsthm}
\usepackage{amssymb} % for \blacksquare
\newtheorem{theorem}{Theorem}
\newtheorem{definition}{Definition}
\newtheorem{lemma}{Lemma}
\newtheorem{statement}{Statement}

\usepackage[utf8]{inputenc} % allow utf-8 input
\usepackage[T1]{fontenc}    % use 8-bit T1 fonts
\usepackage{hyperref}       % hyperlinks
\usepackage{url}            % simple URL typesetting
\usepackage{booktabs}       % professional-quality tables
\usepackage{subcaption}    % for subfigure environment
\usepackage{amsfonts}       % blackboard math symbols
\usepackage{nicefrac}       % compact symbols for 1/2, etc.
\usepackage{microtype}      % microtypography
\usepackage{lipsum}
\usepackage{fancyhdr}       % header
\usepackage{graphicx}       % graphics
\graphicspath{{media/}}     % organize your images and other figures under media/ folder

\title{Decision Theory for Large Scale Outlier Detection Using Aleatoric Uncertainty: With a Note on Bayesian FDR
\thanks{\textit{\underline{Citation}}: 
\textbf{R, Warnick. Decision Theory for Large Scale Outlier Detection Using Aleatoric Uncertainty With a Note on Bayesian FDR.}} 
}

\author{
  Ryan Warnick \\
  Microsoft Security Research \\            
  Microsoft\\
  Redmond, Washington\\
  ryanwarnick@microsoft.com}

\begin{document}
\maketitle

\begin{abstract}
Aleatoric and Epistemic uncertainty have achieved attention recently in the literature as different sources from which uncertainty can emerge in stochastic modeling. Epistemic being intrinsic or model based notions of uncertainty, and aleatoric being the uncertainty inherent in the data. We propose a novel decision theoretic framework for outlier detection in the context of aleatoric uncertainty; in the context of Bayesian modeling. The model incorporates Bayesian false discovery rate control for multiplicity adjustment, and a new generalization of Bayesian FDR is introduced. The model is applied to simulations based on temporally fluctuating outlier detection where fixing thresholds often results in poor performance due to nonstationarity, and a case study is outlined on a novel cybersecurity detection. Cyberthreat signals are highly nonstationary; giving a credible stress test of the model.
\end{abstract}

% keywords can be removed

\section{Introduction}

Aleatoric and epistemic uncertainty are two characterizations of uncertainty that became prominent in the machine learning community for people working with Bayesian models \cite{NIPS_Kensall_Gal, wimmer2023quantifying}, and both have recently received increased scrutiny in \cite{smith2025rethinkingaleatoricepistemicuncertainty}.Explained simply, epistemic uncertainty is the uncertainty in the model, and aleatoric uncertainty is the uncertainty inherent in the data which is distributed accroding to the model. These two types of uncertainty have interactions with each other, with more uncertainty in the model necessary precluding more uncertainty in the data generating mechanism, and as the data increasing model uncertainty decreasing \cite{NIPS_Kensall_Gal}. The reason this comes up naturally in the Bayesian context is because the model has some inherent uncertainty. Thinking of a parametric model, acknowledging in the Bayesian paradigm our uncertainty in the parameters we get that a distribution over the parameters maps to a distribution over a finite dimensional subspace of all distributions on the data. Thus the two types of uncertainty: our uncertainty in the model, epistemic, and our uncertainty in the data generating mechanism, aleatoric.

This paper seeks to analyze the domain of large scale outlier detection, particularly in the framework where we are modeling a large number of independent time series and outliers are occurring during independently across the full spectrum of time series and at each time step. Section \ref{sec:alea_epis} introduces the concept of Aleatoric and epistemic uncertainty and a simple model which we apply in simulations in Section \ref{sec:simulations} to nonstationary time series data using a lagged Bayesian approach. Reviews of outlier detection methods for general data can be found in \cite{outlier_detection_1,outlier_detection_2}. Forecasting models such as Gaussian processes \cite{Rasmussen2006Gaussian} are frequently used in time series analysis, and notions such as z-scores or credible intervals of the forecasting model used to classify outliers.

Additionally, we seek to introduce a novel decision-theoretic formulation adapted to aleatoric uncertainty in Section \ref{sec:decision theory}, and connect it to a Bayesian FDR \cite{FDR} threshold adjusted decision criteria in Section \ref{sec:data_informed}. Threshold selection criteria to control for multiplicity were introduced as early as \cite{Benjamini1995}. Theorems are proven in Section \ref{subsec:note_about_BFDR} showing that for particular classes of threshold selection criteria, the expectation of the decision rule incorporating the threshold selection criteria maps linearly. This connection has been illustrated before by \cite{FDR,Genovese2002FDR}, however a lemma is proposed (Lemma \ref{lemma1}) which opens this framework up to a broader class of problems. This shows that there exists a class threshold selection criteria for which the Bayes optimal rule for which a Bayes optimal decision on a loss incorporating the frequentist analog of the threshold selection rule is the same as the threshold selection criteria. 

We illustrate this for the Bayesian FDR, and additionally, in Section \ref{sec:additional_note} introduced a novel generalization of the Bayesian FDR with parameter $a\in[0,\infty)$; $BFDR(q;a)$. This is defined and illustrated in Sections \ref{subsec:definition} and \ref{subsec:illustration}, and a connection with $L_p$ norm regression \cite{LP_Norm} and the duality between the parameter $a$ and $p$ is proven in Theorem \ref{theorem2}. A comment is added discussing the fact that this threshold selection criteria does not satisfy the assumptions of the previous lemma, but permits a smooth relaxation of threshold dependent Bayes optimal decision rules. A vectorization scheme is proposed improving stability of computation, at the cost of a larger memory requirement, for the generalized $BFDR(q;a)$ in Theorem \ref{theorem3} in Section \ref{subsection:tensorized}, and simulations are done illustrating improved stability across a range of grid resolutions for candidate $\eta$ values in $\eta = BFDR(q;a)$ in the Appendix Section \ref{appendix:clock_time}. We note that this vectorization scheme applies to the original Bayesian FDR as the adjusted Bayesian FDR is a generalization and contains the Bayesian FDR as a special case for $a=1$.

Additionally, in Section \ref{sec:nonparametric} we note that aleatoric uncertainty for a certain class of Gibbs-type nonparametric priors \cite{De_Blasi_2015} is frequently available in closed form, and note that the architecture presented in the rest of the article is applicable in that domain as well. A simulation study is illustrated validating the performance of the BFDR versus fixed values of $\eta$ in a highly nonstationary and demanding environment in Section \ref{sec:simulations}, showing improved performance in precision and recall across a wide range of thresholds $\eta =BFDR(q;a)$ for $a=2$ and providing an adequate stress test of the aleatoric model. A case study on cybersecurity signals is outlined in Section \ref{sec:case_study}. Finally, Section \ref{sec:future_work} outlines considerations for future development, followed with concluding remarks in Section \ref{sec:conclusion}

\section{Illustration of Aleatoric and Epistemic Uncertainty with a Simple Model}
\label{sec:alea_epis}
To see why the concept of aleatoric uncertainty is useful for large scale outlier detection let us analyze a simple model. Suppose the data are assumed to follow $X_i \overset{iid}{\sim}\textrm{N}(\mu,\sigma^2)$, and we place a Normal-Inverse-Gamma prior, $\textrm{NIG}(\mu_0, \nu, \alpha,\beta)$, on the parameters $ \mu, \sigma^2$. Then we have that, given data $X_1,...,X_n$:

 \begin{equation}
  \mu, \sigma^2 | \{X_i\}_{i=1}^n \sim \textrm{NIG}(\frac{\nu\mu_0 + n\overline{X}}{\nu + n}, \nu + n, \alpha + \frac{n}{2}, \beta + \frac{\sum_{i=1}^n (X_i - \overline{X})^2}{2} + \frac{n \nu }{\nu + n}\frac{(\overline{X} - \mu_0)^2}{2}) 
\end{equation}

These parameters are fairly complicated, so denote the parameters:

\begin{equation}
\mu_0^*, \nu^*, \alpha^*, \beta^*, \textrm{ such that }  \mu, \sigma^2 | \{X_i\}_{i=1}^n \sim \textrm{NIG}(\mu_0^*, \nu^*, \alpha^*, \beta^*). 
\end{equation}

Referring to the point made previously, we have a distribution over $ \mu$ and $ \sigma^2$ which is maps to a two-dimensional subspace of all distributions over $X$. This is the epistemic uncertainty. However, if we use the posterior predictive distribution of $ X | X_1,...,X_n$ we get the aleatoric uncertainty:

\begin{equation}
 P(X | X_1,...,X_n) = \int_{\mu \in \mathbb{R}}\int_{\sigma \in \mathbb{R}^+}P(X | \mu, \sigma^2)P(\mu,\sigma^2| X_1,....,X_n)d\sigma^2d\mu
\end{equation}

which luckily for this simple model has a closed form solution:
\begin{equation}
X | X_1,...,X_n \sim \textrm{t}_{2\alpha^*}(\mu_0^*,\frac{\beta^*(\nu^* + 1)}{\nu^*\alpha^*})
\end{equation}

Where the distribution is a non-centered student-t distribution, with $ 2\alpha^*$ degrees of freedom, location parameter $ \mu_0^*$, and scale parameter $\frac{\beta^*(\nu^* + 1)}{\nu^*\alpha^*}$.

\section{Where Decision Theory Comes In}
\label{sec:decision theory}
Where this idea of uncertainty in the data being inherited from the model uncertainty becomes useful is in outlier detection. Say we observe an observation $ X^*$ and we want to know if it's an outlier or not, we can look at $P(X^*> X | X_1,...,X_n)$ and select a cutoff to classify the observed value as an outlier if this probability is too large. The problem becomes choosing this cutoff.

Now to incorporate some ideas from decision theory, suppose we observed $ \{X_{ij}\}_{i \in \{1,...,n\},j\in \{1,...,m\}}$ and want to observe if set of observations $\{X^*_{(n+1)j}\}_{j \in \{1,...,m\}}$ has outliers. If agree with the philosophical leap that we can make decisions based on unobserved predictive values, $ \{X_{(n+1)j}\}_{j\in \{1,...,m\}}$, then our goal is to make a decision $ \delta_j \in \{0,1\}$; where $ \delta_j = 1$ means that $X^*_{(n+1)j}$ was an outlier, and $ \delta_j =0$ means it was not. We can construct loss functions similar in spirit to the following:

\begin{equation}
 L(\delta, X^*, X) = - \sum_{j=1}^m \delta_j \textrm{I}(X^*_{(n+1)j} > X_{(n+1)j} ) + c_1\sum_{j=1}^m (1-\delta_j)\textrm{I}(X^*_{(n+1)j} > X_{(n+1)j} ) + c_2D
\end{equation}

 Here $\textrm{I}(\circ)$ denotes an indicator function, a function returning $1$ if the statement $\circ$ is true and $0$ otherwise, and $D = \sum_{j=1}^m \delta_j$ is a penalty for total discoveries to control for multiplicity. This loss function follows closely in spirit to the loss function in Section 4 of \cite{FDR}, $L(m,\delta, x) = \sum_j\delta_jm_j + k\sum_j (1-\delta_j)m_j + cD$, where instead of a continuous variables $m_j$ we have binary flags, but still being in concordance directionally. E.g. as $X^*_{(n+1)j} \rightarrow \infty$ , then $\textrm{I}(X^*_{(n+1)j} > X_{(n+1)j} )$ monotonically increases to 1.

This loss function provides an incentive for true positives in the first addendum, $ - \sum_{j=1}^m \delta_j \textrm{I}(X^*_{(n+1)j} > X_{(n+1)j} )$, analogous to $\sum_j\delta_jm_j$, an adjustable penalty for false negatives in the second addendum, $c_1\sum_{j=1}^m (1-\delta_j)\textrm{I}(X^*_{(n+1)j} > X_{(n+1)j} )$ which is analogous to $\sum_j (1-\delta_j)m_j$, and a penalty for total discoveries that can be made stronger or weaker, $ c_2D$; which is the same as in the loss in \cite{FDR}.

In typical decision theory we use the posterior expected loss and minimize it, but in this context we're working with aleatoric uncertainty, again under the assumption of the philosophical leap taken earlier that we can work with unobserved predictive values. This means that we are taking the posterior predictive expected loss. This gives us:

\begin{align}
\mathbb{E}_{X|X_1,...,X_n}[L(\delta, X^*, X)] =  - \sum_{j=1}^m \delta_j P(X^*_{(n+1)j} > X_{(n+1)j} | X_{1j},...,X_{nj}) + &\\ \notag c_1\sum_{j=1}^m (1-\delta_j) P(X^*_{(n+1)j} > X_{(n+1)j} | X_{1j},...,X_{nj}) + c_2D 
\end{align}

Denoting by $ r_j $ the value $ P(X^*_{(n+1)j} > X_{(n+1)j} | X_{1j},...,X_{nj})$, we readjust the posterior predictive expected loss to the following:

\begin{equation} \mathbb{E}_{X|X_1,...,X_n}[L(\delta, X^*, X)] =     -\sum_{j=1}^m \delta_j r_j  + c_1\sum_{j=1}^m (1-\delta_j)r_j + c_2D
\end{equation}

For which the solution, minimizing with respect to $ \{\delta_j\}_{j\in\{1,...,m\}}$, is:

 \begin{equation}
 \delta_j = \textrm{I}(r_j > \frac{c_2}{1+c_1})
 \end{equation}

This follows exactly in line with the result of the expected loss minimization in Section 4 of \cite{FDR}, $\textrm{I}(\mathbb{E}[m_j| Y] > \frac{c}{1+k})$, just using the posterior predictive to calculate probabilities for outliers from the indicator functions, but with the same lower bound on the minimizer of the expected loss.

\section{Incorporating Multiplicity Control Into the Loss and Data Informed Loss Functions}

\label{sec:data_informed}

This means that the optimal decision for this loss function, indicator functions of outliers, is constructed out of some lower bound on $ r_j$. One way to select this lower bound instead of fine-tuning the parameters $c_1$ and $c_2$ is to optimize the Bayesian FDR outlined in \cite{FDR}. The posterior predictive CDF values, $ r_j$, provide evidence in favor of testing the hypothesis $ H_0: X^*_{(n+1)j} > X_{(n+1)j}$, for unobserved future predictive values $ X_{(n+1)}$, versus the alternative hypothesis $ H_1: X^*_{(n+1)j} \leq X_{(n+1)j} $. Thus, the threshold could alternatively be set with multiplicity control by finding, for some fixed $ q\in [0,1]$, the threshold would be the largest such $\eta$ to satisfy the following equation:

\begin{equation}
BFDR(q) = \argmax_\eta \frac{\sum_{j=1}^m r_j(I_{(r_j \leq \eta)})}{\sum_{j=1}^m I_{(r_j\leq\eta)}} < q
\end{equation}

This provides a multiplicity control in the correction and offers a less controlled way to manage the false discoveries than tinkering directly with $c_1$ and $ c_2$. 

Let's now examine, in the context of the previous decision theoretic model, what such a Bayesian FDR dictated threshold means in terms of the values of penalties on multiplicity and false negatives. We set up the following equation and solve for $ c_1$ and $ c_2$:

\begin{equation}
 BFDR(c_2/(1+c_1)) = \argmax_\eta \frac{\sum_{j=1}^m r_j(I_{(r_j \leq c_2/(1+c_1))})}{\sum_{j=1}^m I_{(r_j\leq c_2/(1+c_1))}} < q
\end{equation}

Rearranging a bit we get: $\sum_{j=1}^m (r_j-q) I_{(r_j \leq c_2/(1+c_1))}< 0$. In other words, the largest such pairing of $ c_1$ and $ c_2$ such that the magnitude of the sum of all $ \{ r_j - q: r_j > q \land r_j \leq c_2/(1+c_1)\}$ is less than the magnitude of the sum of all $ \{ r_j - q: r_j \leq q \land r_j \leq c_2/(1+c_1)\}$ (which is going to be less than or equal to 0 because $ r_j \leq q$).

As $ c_2/(1+c_1)$ gets smaller, the number of elements in the first set decreases more rapidly than the number of elements in the second set. This is because the second logical expression makes the first less likely to be true. Inversely, the same is true in the opposite context in the opposite direction, as $ c_2/(1+c_1)$ gets larger the number of elements increases more rapidly in the first set. An equilibrium needs to be find as the largest such $ c_2/(1+c_1)$ such that the statement is true.

Note that if $ c_2/(1+c_1) = \eta$ for some $ \eta : BFDR(\eta) < q$, then we have that $ c_2 = (1+c_1)\eta$ and vice versa $ c_1 = \frac{c_2}{\eta} - 1$. Remembering that $ c_1$ is the penalty for false negatives, this gives us the linear relationship dictated by the BFDR threshold between the penalty for false negatives and the penalty for total discoveries. We have that $ BFDR^{-1}(q; r_1,...,r_n)$ is a monotonically (not strictly) decreasing function; for which the rate of increase depends on $ \{r_1,...,r_n\}$

Thus, holding the penalty on false negatives, $ c_1$, fixed, we get $ c_2 = (1+c_1)BFDR^{-1}(q)$ is a where the right hand side is a composition of a linear and monotonic function, this monotonicity as a function of $ q$. Note that as $ q$ decreases $ c_2$ increases monotonically, and as $ c_1$ increases $ c_2$ increases linearly. This is different from holding $ c_2$, the penalty for total discoveries fixed, where an increase in $ c_2$ causes a linear decrease in $ c_1$.

\begin{equation}
L(\delta, X^*, X) = - \sum_{j=1}^m \delta_j \textrm{I}(X^*_{(n+1)j} > X_{(n+1)j} ) + c_1\sum_{j=1}^m (1-\delta_j)\textrm{I}(X^*_{(n+1)j} > X_{(n+1)j} ) +(c_1 + 1)BFDR^{-1}(q)D
\end{equation}

From this point it becomes clear; for appropriately specified loss functions and thresholds, one can reverse engineer decision criteria to give a loss function for which a particular threshold selection criteria (whatever type, whatever appropriately specified  loss function) coincides. 

However, this is somewhat circular logic, since the Bayesian FDR is necessarily Bayesian and is taken and evaluated after the expectation is taken. Still we believe there is an interesting potential to proceed forward with this train of thought. In essence, it the formulations says that we have a data-dependent loss function. In other words, if you use the Bayesian FDR a posteriori to select a cutoff, it would have equated to using this decision criteria on that data.  Section \ref{subsec:note_about_BFDR} gives situations in which this asymmetry is broken and the expected loss solution is the BFDR threshold.

\subsection{Note About the Bayesian FDR in this Context}
\label{subsec:note_about_BFDR}
Assume in a typical Bayesian model that we have  $ \{\gamma_j\}_{j=1}^m | Y \in \{0,1\}^m$ which is some model.  Let $ r_j = \mathbb{E}_{\gamma_j | Y}[\gamma_j]$.

Examine the following loss:

\begin{equation}
\label{eq:bfdr_loss}
L(\delta, \gamma, X) = - \sum_{j=1}^m \delta_j \gamma_j + c_1\sum_{j=1}^m (1-\delta_j)\gamma_j +FDR^{-1}(q; \gamma)D
\end{equation}

Where $ q = FDR(\eta; \gamma)$ is the false discovery rate (which is monotonic in $ q$ because it can only be 0 or 1 in this scenario; prior to taking the expectation). Note also that $ FDR^{-1}(q; \gamma)$ is monotonic in each $ \gamma_j$. We prove a small theorem here to show that $ \mathbb{E}_{\gamma | Y}\textrm{FDR}^{-1}(q;\gamma)] = BFDR^{-1}(q; r)$, but rely on the following lemma proved in the appendix:
\newline
\begin{lemma}
\label{lemma1}
Suppose $f: \{\delta\}_{j=1}^m \in \{0,1\}^m \rightarrow q$ is a random function that is non-negative and piecewise-constant, monotonic in $\delta_j$ for some direction for each $\delta_j$; with induced randomness in $f$ coming from another random variable $\gamma$ which is a binary vector $\gamma_j \overset{ind.}{\sim} \sim\textrm{Bern}(r_j)$. Assume additionaly that $f$ is almost surely monotonic (in either direction) as a function of each $\gamma_j$. Suppose $g: q \rightarrow \eta = \argmax_\eta \{f(\eta): f(\eta) < q\}$.Then:

\begin{enumerate}
\item $\mathbb{E}\gamma[g(q;\gamma)] = \mathbb{E}\gamma[g; r_j](q) =  \argmax_\delta \{\mathbb{E}[f; r_j](q) : \mathbb{E}[f; r_j](q) < q\}]$
\item $\mathbb{E}\gamma[g; r_j]$ is interconnected with the monotonicity in $\gamma$ and $\delta$. If $f$ is monotone increasing (or decreasing) with respect to each $\gamma_j$, and monotone increasing (or decreasing) in the same direction with respect to each $\delta_j$, then the monotonicity of $\mathbb{E}\gamma[g; r_j]$ with respect to $\delta_j$ is the inverse of the monotonicity in $\delta_j$ before the expectation is taken. If they are in the opposite direction the monotonicity in $\delta_j$ is the same after the expectation is taken.  
\end{enumerate}

\begin{proof}
Proof in Appendex Section \ref{prooflemma1}
\end{proof}
\end{lemma}

\begin{theorem}
\label{theorem1}
Statement: For $ \gamma_j| Y \overset{ind.}{\sim}\textrm{Bern}(r_j), \hspace{2mm} j\in\{1,...,m\}$, and $FDR(\delta) = \frac{\sum_{j=1}^m\delta_j(1-\gamma_j)}{\sum_{j=1}^m \delta_j}$, we have the following:

\begin{equation}
\mathbb{E}_{\gamma| Y }[FDR^{-1}; r_j](q) = BFDR^{-1}(q; r_j)
\end{equation}

Additionally $BFDR^{-1}(q;r_j)$ is monotonically increasing as a function of each individual $r_j$.
\end{theorem}

\begin{proof}
Note that $FDR(\delta ; \gamma_j) = \frac{\sum_{j=1}^m\delta_j(1-\gamma_j)}{\sum_{j=1}^m \delta_j}$ is a random function of $\gamma$, which is non-negative and piecewise constant with respect to both $\delta_j$ and $\gamma_j$. Note that $\gamma|Y \overset{ind.}{\sim} \textrm{Bern}(r_j)$. Note that the value is monotonically decreasing as a function of each $\gamma_j$, and becoming monotonically closer to $1$ as a function $\delta_j$ for fixed $\gamma_j$. 

Note that for fixed $q$ the $BFDR(q; r_j)$ in this context is:

\begin{equation}
\argmax_\delta\{\mathbb{E}\gamma[FDR(\delta;\gamma); r_j] : \mathbb{E}[f; r_j](q) < q\}
\end{equation}

This satisfies the requirements for Lemma \ref{lemma1}.

However, $\mathbb{E}\gamma[FDR(\delta;\gamma); r_j] = \frac{\sum_{j=1}^m \delta_j (1-r_j)}{\sum_{j=1}^m \delta_j}$ by linearity of expectation \cite{FDR}. This gives us that the previous equation is $\argmax_\delta \{\mathbb{E}\gamma[FDR(\delta;\gamma); r_j] : \mathbb{E}[f; r_j](q) < q\} = \argmax_\delta \{\frac{\sum_{j=1}^m \delta_j (1-r_j)}{\sum_{j=1}^m \delta_j} : \frac{\sum_{j=1}^m \delta_j (1-r_j)}{\sum_{j=1}^m \delta_j} < q\}$. This is the inverse of Bayesian FDR.

$ \blacksquare$
\end{proof}

Note also that in the previously specified loss in equation \ref{eq:bfdr_loss}, the $FDR(\delta,\gamma)$ denominator $D$ divides out with the multiplicity control in the loss, givins us $\sum_j \delta_j(1-\gamma)$, which still satisfies the requirements of Lemma \ref{lemma1}. This allows us to take the expectation and then multiply by a complex form of $1$ ($\frac{D}{D}$) to get the expectation of Equation \ref{eq:bfdr_loss} is :

\begin{equation}
\label{eq:expected_bfdr_loss}
\mathbb{E}{X|X_1,....,X_n}L(\omega, \gamma, X) = - \sum_{j=1}^m \delta_j r_j + c_1\sum_{j=1}^m (1-\delta_j)r_j +BFDR^{-1}(q;r)D
\end{equation}

Which has solution $\delta_j = I_{(r_j > \frac{BFDR^{-1}(q;r)}{1+c_1})}$. 

\section{Additional Note on Bayesian FDR}
\label{sec:additional_note}
\cite{BFDR_Model_Misspecification} previously considered the performance of BFDR under various types of model misspecification. We introduce a modified $ BFDR(q;a)$ here to account for certain types of model misspecification.

\subsection{Definition}
\label{subsec:definition}
\begin{definition}

\begin{equation} BFDR(\eta; a) =\argmin_q \{\sum_{j=1}^m\textrm{sign}(r_j - q)|r_j - q|^a I_{r_j \leq \eta}:\sum_{j=1}^m\textrm{sign}(r_j - q)|r_j - q|^a I_{r_j \leq \eta} < 0\}
\end{equation}

For $ a \in \mathbf{R}^+\cup{0}$.
\end{definition}

\subsection{Illustration}
\label{subsec:illustration}
We have that:

\begin{equation}
\eta =  \argmax_\eta \{\sum_{j=1}^m \textrm{sign}(r_j-q) |(r_j - q)|I_{(r_j \leq \eta)} : \sum_{j=1}^m \textrm{sign}(r_j-q) |(r_j - q)|I_{(r_j \leq \eta)} < 0 \}
\end{equation}

Using the sets outlined 2 sections ago we can see that the previous statement is true. 

We note that this expression for $ BFDR^{-1}(q)$ is piecewise constant and increasing with respect to latex $q$, and that for a fixed $ \eta$ the BFDR is the minimum such value of $ q$ that this expression is true. (This is because the BFDR is satisfied by more than one such $ q$, but the minimum value is the true BFDR). This gives us that

\begin{equation}
q = \argmin_q \{\sum_{j=1}^m\textrm{sign}(r_j - q) |(r_j - q)I_{r_j \leq \eta}| : \sum_{j=1}^m\textrm{sign}(r_j - q) |(r_j - q)I_{r_j \leq \eta}|<0\}
\end{equation}

Note that, $ r_j - q \in [-1,1]$, giving us that $ |r_j - q| \in [0,1]$, giving us that $ |r_j - q|^a $ is going to similarly be in [0,1] for any $ a \in \mathbf{R}^+\cup{0}$. This allows us to modify the previous expression to be:

\begin{equation}
q = \argmin_q \{\sum_{j=1}^m\textrm{sign}(r_j - q) |r_j - q|^a I_{r_j \leq \eta} : \sum_{j=1}^m\textrm{sign}(r_j - q) |r_j - q|^a I_{r_j \leq \eta}  < 0\}
\end{equation}

Note that as $ a \rightarrow \infty $ the differences of the $r_j$ are further seperated relative to $q$, and for values of $a \rightarrow 0$ going towards $ 0$ extreme values are brought closer together. It should be clear that we can reverse engineer this to get an appropriate $\eta$ for $BFDR(\eta; a) < q$.

Another interesting point to consider is that $ \sum_{j=1}^m \textrm{sign}(r_j - q)|r_j - q|^aI_{r_j < \eta} = \frac{d}{dq}\sum_{j=1}^m |r_j - q|^{a+1}I_{r_j \leq \eta}$. Giving us that our solution is the point at which the rate of change of the conditional median is maximized. The solution is the minimizer of the change in rate of the conditional absolute $a+1$ non-central moment of the $ \{r_j\}_{j=1}^m$. E.g. with respect to the non-centrality parameter:

\begin{theorem}
\label{theorem2}
Let $ R \sim \frac{1}{m}\sum_{j=1}^m \delta_{r_j}(\circ)$ and we have the following:

\begin{equation} 
\forall a \geq 0, \hspace{2mm} BDFR(\eta; a) = \argmin_{q} \{\frac{d}{dq}\mathbb{E}_R[|r - q|^{a+1}| R \leq \eta] : \frac{d}{dq}\mathbb{E}_R[|r - q|^{a+1}| R \leq \eta] < 0\}
\end{equation}

Which is the point which minimizes the change in the $a+1$ non-central absolute moment with respect to the non-centrality parameter.
\end{theorem}

\begin{proof}
The proof follows from applying the dominated convergence theorem to exchange an expectation and a derivative and using laws from basic calculus. There is a $ \frac{1}{a+1}$ term that has to be considered, but this term can be removed from the expectation and derivative, and the minimum is invariant under scalar multiplication/division for scalars greater than 0, so we can remove this term as well. The same applies to the normalizing constant in the distribution of $ R$, $ \frac{1}{m}$.

$ \blacksquare$
\end{proof}

This reduces the estimation of $BFDR$ to an $L_p$ norm regression problem on the $r_j$. \cite{LP_Norm} have previously investigated regression problems under $L_p$ norms, and showed that for non-Gaussian distributions of the data ($\{r_j\}_{j=1}^m$) values of $p$ other than $2$ ($a=1$) give optimal results. They also illustrate ways to optimize based on kurtosis, which could allow the $BFDR(q;a)$ parameter $a$ to be estimated from the kurtosis of the values $\{r_j\}_{j=1}^m$ to achieve optimality. The support of the $\{r_j\}_{j=1}^m\subset [0,1]$ means that the data is strictly not Gaussian, lending credence to possible improvements using the adjusted $BFDR(q;a)$. Note that for $a =0$ we recover the median of the $\{r_j\}_{j=1}^m$ values

This theorem also provides us a way to investigate complexities such as those that arise when $ \{r_j\}_{j=1}^m$ is random and correlated. In other words, conditional on $ R$ the $ \gamma$ are independent, but there are dependencies amongst the $ r_j \sim R$. This would equate to ways in which the data contain colinearities or heteroscedasticity in the $L_p$ norm approach discussed in the previous paragraph.

We should note that this Bayesian FDR corresponds to a specific instantiation of an adjusted $FDR(q;a)$, but that this might not be in concordance with Lemma \ref{lemma1} and satisfying the framework of Section \ref{subsec:note_about_BFDR} and Theorem \ref{theorem1}. This means that using this threshold selection criteria is Bayes optimal for a loss function incorporating $FDR(q;a)$ only if that loss was applied to the data the $BFDR(q;a)$ applied to. However, it would be interesting for future developments to investigate if the proximity of $a$ to $1$ (giving the $BFDR(q)$ without modification; which does map over the loss) allows a kind of smooth relaxation of the threshold based loss function. This is a consideration worthy of future investigation.  

\subsection{Tensorized Adjusted \texorpdfstring{$BFDR(q;a)$}{BFDR(q;a)}}
\label{subsection:tensorized}
Computing the adjusted BFDR by looping over a sequence of values is prohibitively expensive for large data sets, and also limits the capacity to consider a high-resolution grid of values due to performance inefficiencies.

However, it is possible reduce the $BFDR(q;a)$ to tensor operations to greatly reduce the computational bottleneck. Additionally, because of the generality of the adjusted $BFDR(q;a)$ in containing the original $BFDR(q)$, this is beneficial for practicioners generally.

To reiterate:

\begin{equation}
\eta =  \argmax_{\eta\in \vec{K}} \{\sum_{j=1}^m \textrm{sign}(r_j-q) |(r_j - q)|^aI_{(r_j \leq \eta)} : \sum_{j=1}^m \textrm{sign}(r_j-q) |(r_j - q)|^aI_{(r_j \leq \eta)} < 0 \}
\end{equation}

where $\vec{K} = [0, \frac{1}{k}, \frac{2}{k},....,1]$ is some grid with resolution $k\in \mathbb{N}$, is difficult to compute using a loop. This can be greatly improved using vectorization.

\begin{theorem}

\label{theorem3}
For:

\begin{equation}
\eta =  \argmax_{\eta\in \vec{K}} \{\sum_{j=1}^m \textrm{sign}(r_j-q) |(r_j - q)|^aI_{(r_j \leq \eta)} : \sum_{j=1}^m \textrm{sign}(r_j-q) |(r_j - q)|^aI_{(r_j \leq \eta)} < 0 \}
\end{equation}

where $\vec{K} = [0, \frac{1}{k}, \frac{2}{k},....,1]$ is some grid with resolution $k\in \mathbb{N}$, is difficult to compute using a loop. This can be greatly improved using vectorization.

Let $\vec{r} = \begin{pmatrix}
r_1 \\
r_2 \\
\vdots \\
r_m\end{pmatrix}$ and let $R= \vec{r}\otimes \vec{1}_{(k+1)}^T$. We construct an indicator matrix $\Psi$ corresponding to the following:

\begin{equation}
\Psi = [I(R_{jl} <(\vec{K}\otimes \vec{1}_{m}^T)^T_{jl})]_{jl}
\end{equation}

This matrix is of dimension $m\times (k+1)$. The $j$ index corresponds to the anomaly scores and the $l$ index corresponds to the grid. This can also be computing quickly through vectorization.

Then we have that the $BFDR(q;a)$ can be computed as follows:

\begin{equation}
\eta = \frac{\textrm{\textit{max index}}_l \{\vec{1}_m^T[\textrm{sign}(R\odot \Psi - q\times\Psi)\odot |R\odot \Psi - q\times\Psi|^a] < 0 \}}{k}
\end{equation}

Where the sign function, absolute value, and exponentiation are taken elementwise.

\end{theorem}

Note that the $\vec{1}_m^T$ term in $\vec{1}_m^T[\textrm{sign}(R\otimes \Psi - q\times\Psi)\otimes (R\otimes \Psi - q\times\Psi)^a]$ is for the vector $\times$ matrix product to sum over the axis of the anomaly scores, and $\sign(\circ)$ is a function which returns an entry of $1$ in the matrix if the value is positiv and $-1$ if the value is negative. The exponential is performed element-wise.

\begin{proof}
Proof in Appendix Section \ref{prooftheorem3}
\end{proof}

Testing on one of our simulations. We had $1500$ replications on $m=1000$ for $a=2$ and $k=10000$, and we selected $q=.2$. For the tensorized form the total clock time was $2$ minutes and $53$ seconds, mean of click time across all replications was $.110831$ seconds, and standard deviations of the clock times was $.00272$ seconds. 

\begin{figure}
        \caption{Histogram of clock times of tensorized and looped $BFDR(q;a)$; for 1500 replications, $k=10000$, $q=.2$, and $a=2$. Histogram for looped is shown in red, and for vectorized is shown in blue.}
        \label{fig:clocktime_distribution}
        \centering
        \includegraphics[width=0.7\textwidth]{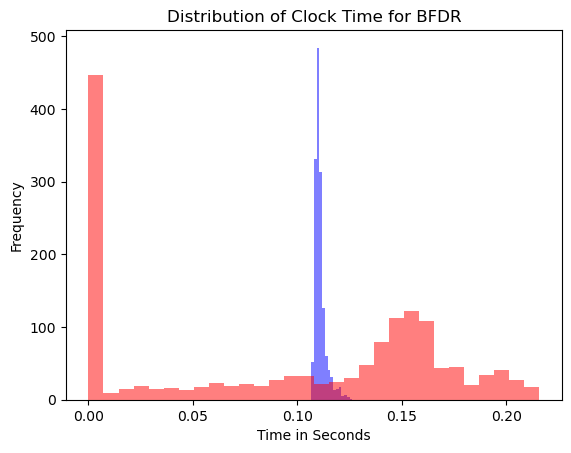}
        \caption{Histogram of clock times of tensorized and looped $BFDR(q;a)$; for 1500 replications, $k=10000$, $q=.2$, and $a=2$. Histogram for looped is shown in red, and for vectorized is shown in blue.}
\end{figure}

The looped clock time got a better total clock time, but had much higher variance in individual clock times across replications. For the same set of replications the total clock time was $2$ minutes and $20.7$ seconds, mean clock time was $.093832$ seconds, and the standard deviation of clock times was $.07293$ seconds. Figure \ref{fig:clocktime_distribution} illustrates this well. The tendency towards low clock times with the looping might be that the loop iterates over $K$ from $1\rightarrow 0$ in that order, and potentially the appropriate $BFDR(q;a)$ is discovered at a high value; close to 1. This might create difficulties in situations where the $BFDR(q;a)$ is repeatedly very small.

Additional figures showing performance of the tensorized versus looped $BFDR(q;a)$ can be found in the Appendix Section \ref{appendix:clock_time}.

\section{Nonparametric Models}
\label{sec:nonparametric}
It's possible to get more abstract with the posterior predictive expection for outlier detection. For example, suppose we don't want to use a parametric model, but instead some nonparametric approach. Taking Gibbs-type priors as an example \cite{De_Blasi_2015}, and specifically the Dirichlet process:

\begin{align}
X_i& \overset{\textrm{iid}}{\sim} G\\
\notag G &\sim\textrm{DP}(\alpha, G_0)
\end{align}

With $\alpha >0$ and $G_0$ a probability measure with support of the data. We get that the posterior predictive is:

\begin{equation}
\label{eq:DP_pred}
X | X_1,...,X_n \sim \frac{\alpha}{\alpha +n} G_0(\circ) + \frac{1}{\alpha+n}\sum_{i=1}^n \xi_{X_i}(\circ)
\end{equation}

where $ \xi_x(\circ)$ denotes a point mass measure at $ x$ \cite{T&CDP_Jara}. This can be used to make outlier decisions in a decision theoretic model.

More generally, a Gibbs-type prior is a species sampling models which constitute a generalizing class of the DP and other associated non-parametric priors, such as the Pitman-Yor Process \cite{pitman97}; $\textrm{PY}(\alpha,\theta, G_0)$. So far, many types of predictive distributions for Gibbs-type priors have been classified \cite{Arbel_2020, De_Blasi_2015}, and \cite{Arbel_2020} have shown desirable features across a broad class of Gibbs-type processes as the lag ($n$ in $\{X_1,...,X_n\}$) increases, approaching the desirable properties of the Pitman-Yor Process predictive distribution, which is as follows:

For $\{\sigma \in [0,1]$ and $\theta > -\sigma\}$ $\lor$ $\{\sigma < 0$ and $\theta = m|\sigma|$ for some positive integer $m\in \mathbb{N}\}$, and $k$ being the number of existing ties in $\{X'_1,...X'_k\}= \textrm{Unique}(\{X_1,...,X_n\})$:

\begin{align}
\label{eq:PY_pred}
X_i &\overset{\textrm{iid}}{\sim} G\\
\notag G &\sim \textrm{PY}(\alpha, \theta, G_0)
\end{align}

We have that:
\begin{equation}
X^* | X_1,...,X_n \sim \frac{\theta + \sigma k}{\theta +n}G_0(\circ) +\sum_{i=1}^k\frac{n_k - \sigma}{\theta + n}\xi_{X'_i}(\circ)
\end{equation}

Where $n_k$ is the count of values in $\{X_1,...,X_n\} $ which are equal to $X'_k$. This type of clustering behavior is useful in situations where the data space is truly discrete (such as counts of telemetry in cybersecurity contexts), and a distribution such as a Negative Binomial or Poisson can be specified as $G_0$.

Additionally, in situations where a nonparametric approach is desirable, but the parameterization would like to be left more unrestricted, \cite{De_Blasi_2015}
show that for mixtures of dirichlet processes:

\begin{align}
X_i& \overset{\textrm{iid}}{\sim} G\\
\notag G &\sim\textrm{DP}(\alpha, G_0)\\
\notag \alpha &\sim H
\end{align}

where $H$ is a measure with support $\mathbb{R}^+$, we have the following:

For $P(X^*= \hspace{2mm} \textrm{"new"} \hspace{2mm} | X_1,...,X_n) = b_n^{(k)} = \int_{\mathbb{R}^+} \frac{\theta^k}{(\theta)_n} dH(\theta)$, where $(\theta)_n$ is the n'th ascending factorial.

This shows that mixtures of Dirichlet Processes depend on the number of matches as well; similar in spirit to the Pitman-Yor Process. For a detailed description of predictive distributions across a broad class of Gibbs-type priors refer to \cite{lancelot_james}.

\section{Simulations}
\label{sec:simulations}

In nonstationary multivariate time series, detecting outliers is difficult because the signal to noise ratio evolves over the course of the experiment. Additionally, at every time step, the number of elements of the full set of series that could be parsimoniously detected as outliers might fluctuate wildly, thus it is of critical importance to perform multiplicity control not globally across the full signals across all time steps, but instead for the full set of signals at each time step. 

\begin{figure}
    \centering
    \begin{subfigure}[t]{0.45\textwidth}
        \centering
        \includegraphics[width=\linewidth]{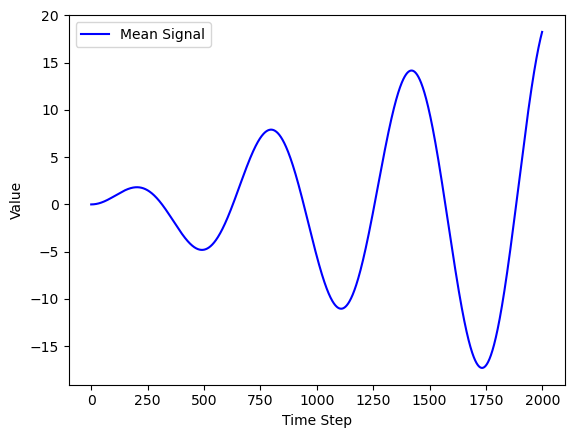} 
        \caption{The mean of the signal around which the data and outliers vary.} \label{fig:mean_signal}
    \end{subfigure}
    \hfill
    \begin{subfigure}[t]{0.45\textwidth}
        \centering
        \includegraphics[width=\linewidth]{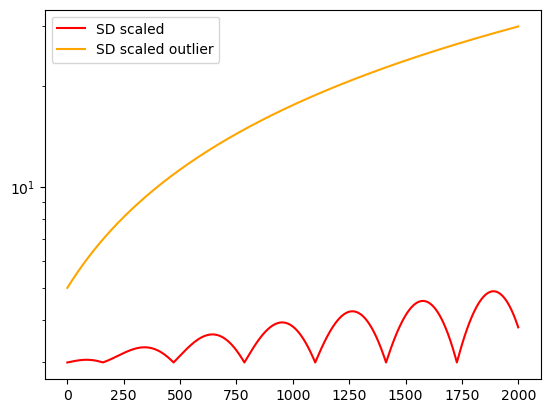} 
        \caption{Standard deviation of outliers and inliers. Outliers are sampled as having both the sum of a random variated generated from $\textrm{N}(\mu_t,\sigma_t^{inlier}) + \textrm{N}(\mu_t,\sigma_t^{outlier})$ to increase the difficulty of detecting the signal.} 
        \label{fig:sd_signal}
    \end{subfigure}

    \vspace{1cm}
    \begin{subfigure}[t]{0.45\textwidth}
        \centering
        \includegraphics[width=\linewidth]{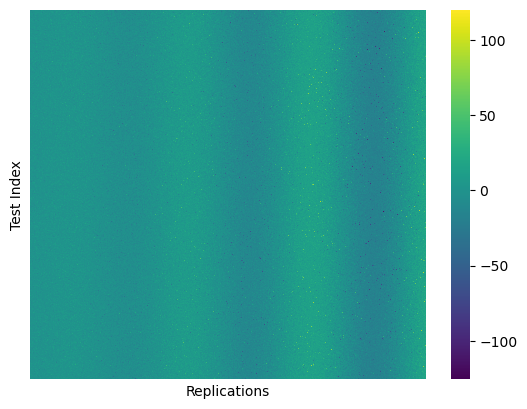} 
        \caption{The data matrix. Colorbar indicates value and the data is oscillating around the mean signal illustrated in Figure \ref{fig:mean_signal}. Standard deviation dictated by Figure \ref{fig:sd_signal}.} \label{fig:data_matrix}
    \end{subfigure}
    \hfill
    \begin{subfigure}[t]{0.45\textwidth}
        \centering
        \includegraphics[width=\linewidth]{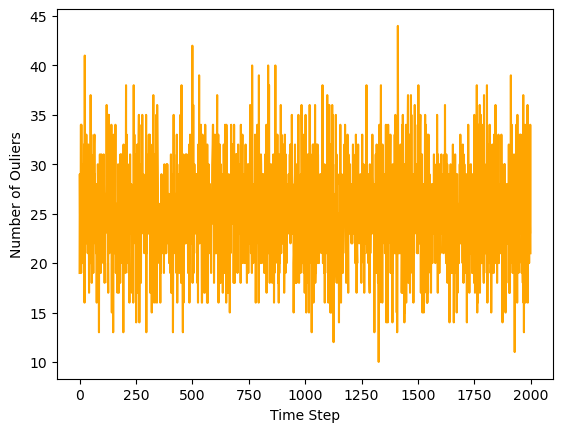} 
        \caption{Number of true outliers per time step. Outliers are selected as 2.5\% of the total set of data, which equates to approximately 25 at every time point, fluctuating roughly between a range of 15-35.} \label{fig:true_outlier_count}
    \end{subfigure}
\end{figure}

We construct a set of simulations outlined in Figures \ref{fig:mean_signal} - \ref{fig:true_outlier_count}. The mean signal is a sinusiodal signal growing in amplitude as a function of the time step, and the outliers are interwoven with the inliers by sampling outliers both with the inlier and outlier standard deviation at each time step. The set $m$ (corresponding to the y axis in figures \ref{fig:mean_signal} - \ref{fig:true_outlier_count}) is $1000$, and the number of time points is $2000$. The data are training on the lagged signal paralellelized over $j\in\{1,...,m\}$ with a lag of 30, giving us an evaluation set of $1000\times 1969$. 

The baseline $\textrm{NIG}(\mu_0,\nu, \alpha,\beta)$ prior is specified to have parameters $\mu_0 = 0$, $\nu = .0001$, $\alpha=.01$ and $\beta = .01$. This is a diffuse NIG prior. The $BFDR(q;a)$ was evaluated across a range of $q \in \{\frac{1}{2^v}\}_{v=1}^{15}$, and for $a=2$. We evaluated the model across a number of performance metrics; including Precision, Recall, Accuracy, and Balanced Accuracy. A description of statistical performance metrics for binary classifiers can be found in \cite{performance_metrics}.

\begin{figure}
    \centering
    \begin{subfigure}[t]{0.45\textwidth}
        \centering
        \includegraphics[width=\linewidth]{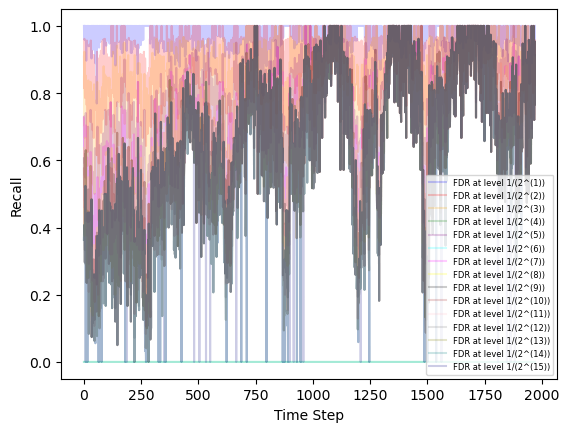} 
        \caption{Recall of the model for $BFDR(q;a)$ for $a=2$ and $q \in \{\frac{1}{2^v}\}_{v=1}^{15}$} \label{fig:recall}
    \end{subfigure}
    \hfill
    \begin{subfigure}[t]{0.45\textwidth}
        \centering
        \includegraphics[width=\linewidth]{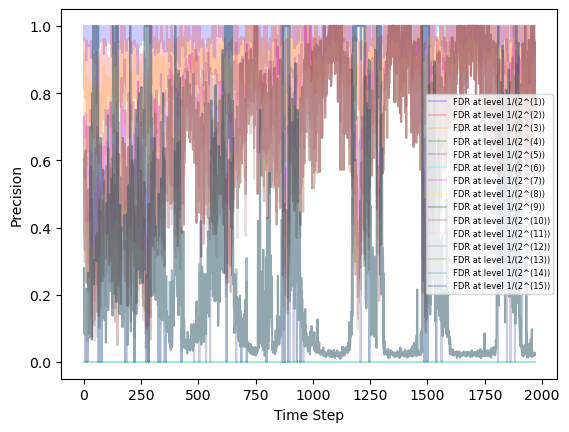} 
        \caption{Precision of the model for $BFDR(q;a)$ for $a=2$ and $q \in \{\frac{1}{2^v}\}_{v=1}^{15}$} \label{fig:precision}
    \end{subfigure}

    \vspace{1cm}
    \begin{subfigure}[t]{0.45\textwidth}
        \centering
        \includegraphics[width=\linewidth]{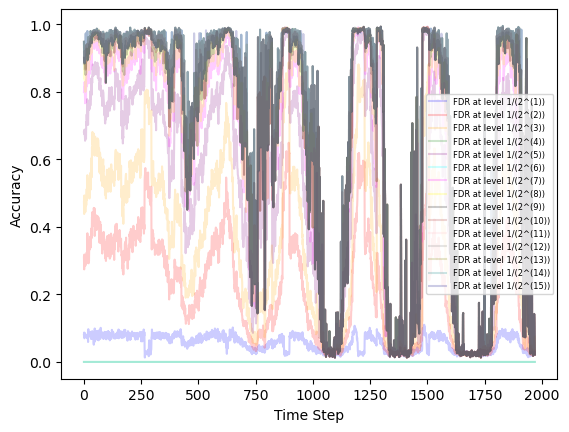} 
        \caption{Accuracy of the model for $BFDR(q;a)$ for $a=2$ and $q \in \{\frac{1}{2^v}\}_{v=1}^{15}$} \label{fig:accuracy}
    \end{subfigure}
    \hfill
    \begin{subfigure}[t]{0.45\textwidth}
        \centering
        \includegraphics[width=\linewidth]{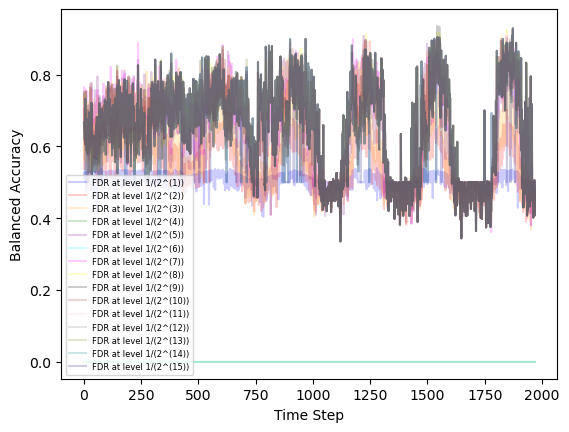} 
        \caption{Balanced Accuracy of the model for $BFDR(q;a)$ for $a=2$ and $q \in \{\frac{1}{2^v}\}_{v=1}^{15}$} \label{fig:balanced_accuracy}
    \end{subfigure}
\end{figure}

We see in Figures \ref{fig:recall} and \ref{fig:precision} that the model gets uniformly high precision and recall for small values of $q$. Accuracy achieves higher performance in low noise paradigms (close to the beginning of the time series, with performance in high noise paradigms being dictated by properties of the noise of the baseline signal ($\sigma_t^\textrm{inlier}$) and the amplitude of the baseline trend ($\mu_t$). Balanced accuracy remains consistently above $50\%$ and sometimes as high as $.929$ later in the series, again dictated by $\mu_t$ and $\sigma_t^\textrm{inlier}$.

\begin{figure}
    \centering
    \begin{subfigure}[t]{0.45\textwidth}
        \centering
        \includegraphics[width=\linewidth]{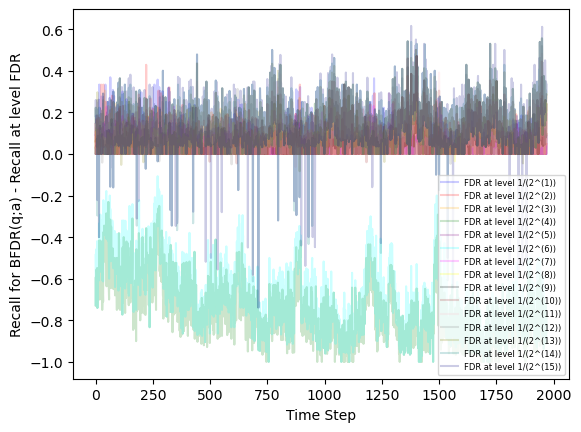} 
        \caption{Recall of the model for $BFDR(q;a)$ for $a=2$, minus recall of the model for $\eta = q$;  $q \in \{\frac{1}{2^v}\}_{v=1}^{15}$}
        \label{fig:recall_diff}
    \end{subfigure}
    \hfill
    \begin{subfigure}[t]{0.45\textwidth}
        \centering
        \includegraphics[width=\linewidth]{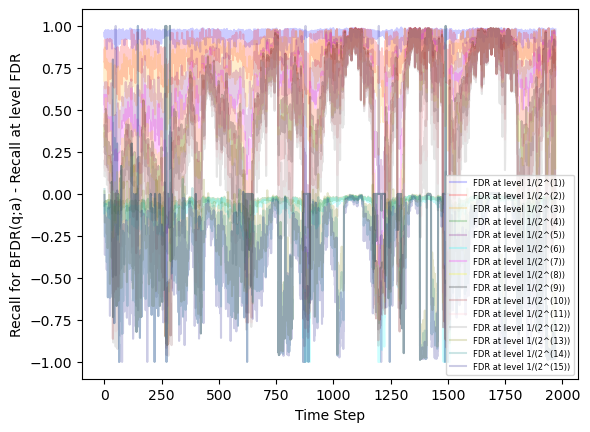} 
        \caption{Precision of the model for $BFDR(q;a)$ for $a=2$, minus precision of the model for $\eta = q$;  $q \in \{\frac{1}{2^v}\}_{v=1}^{15}$} \label{fig:precision_diff}
    \end{subfigure}

    \vspace{1cm}
    \begin{subfigure}[t]{0.45\textwidth}
        \centering
        \includegraphics[width=\linewidth]{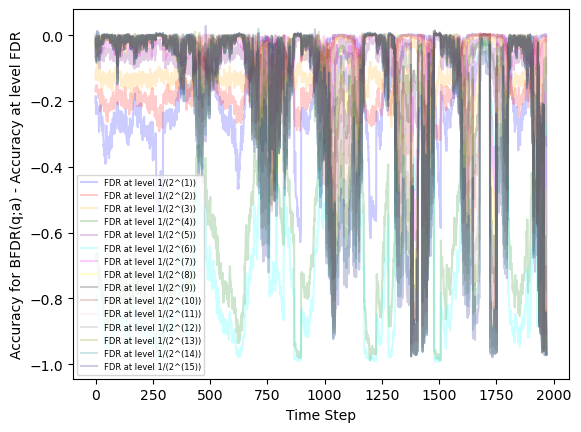} 
        \caption{Accuracy of the model for $BFDR(q;a)$ for $a=2$, minus accuracy of the model for $\eta = q$;  $q \in \{\frac{1}{2^v}\}_{v=1}^{15}$}
        \label{fig:accuracy_diff}
    \end{subfigure}
    \hfill
    \begin{subfigure}[t]{0.45\textwidth}
        \centering
        \includegraphics[width=\linewidth]{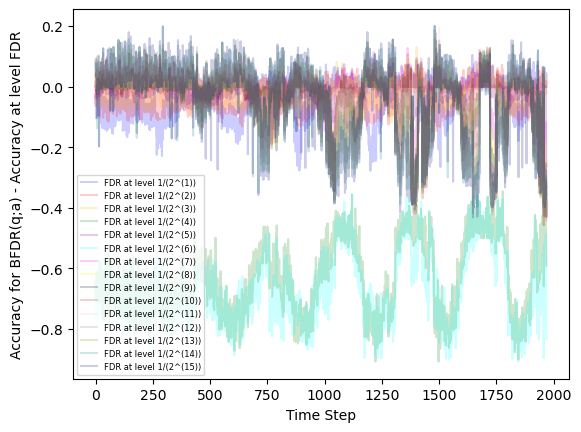} 
        \caption{Balanced Accuracy of the model for $BFDR(q;a)$ for $a=2$, minus balanced of the model for $\eta = q$;  $q \in \left\{\frac{1}{2^v}\right\}_{v=1}^{15}$}
        \label{fig:balanced_accuracy_diff}
    \end{subfigure}
\end{figure}

For comparison, we are interested in Precision and Recall (for which a good description can be found in the chapter on evaluation in \cite{information_retrieval}). As an alternative scenario, assume that we had a typical threshold selection problem of choosing $\eta$ to achieve a controlled FDR at level $q$ for some subset of the samples; and then extrapolate to other time points. The first barrier to this problem is a gap in knowledge, we must first know the truth for the selected time block. This is not realistic in many cybersecurity scenarios. However, assuming we did, we construct simulations for $\eta = \{\frac{1}{2^v}\}_{v=1}^{15}$ and threshold appropriately. This gives us the results illustrated in Figures \ref{fig:recall_diff} - \ref{fig:balanced_accuracy_diff}, where each plot is the difference across the full $1969$ time points of each appropriate metric for the $BFDR(q;a)$ and $\eta = q$ at the same level. Values above 0 show better performance using $BFDR(q;2)$, values below 0 show better performance just in strict thresholding at value $q$.

We can

We can see that in terms of recall, our model does nearly uniformly better for sufficiently small values of $q$, and for precision similarly performs better for sufficiently small values of $q$. Accuracy performs nearly uniformly worse; however, balanced accuracy for sufficiently small values of $q$ performs adequately equal, sometimes performing worse in high noise paradigms as a function of $\mu_t$ and $\sigma_t^{\textrm{inlier}}$. Our application is cybersecurity, where precision and recall are the performance metrics most appropriate for study; as accuracy and retrieval are of utmost importance. These results on precision and recall indicate improvement over fixed thresholding. 

Note also the $(1+c_1)$ term in the threshold in Equation \ref{eq:expected_bfdr_loss} in Section \ref{subsec:note_about_BFDR}. This term is a control for false negatives, and with the result in Theorem \ref{theorem1}, gives us a threshold at each time step of $r_{j_t}>\frac{\eta_t}{1+c_1}$. Increasing this threshold $c_1 \rightarrow \infty$ increases recall, and lowering it $c_1\rightarrow 0$ gives us the results outlined in this section; where it is detected a an outlier if $r_{j_t} > \eta_t$.

\section{Case Study}
\label{sec:case_study}

\section{Possible Extensions and Future Work}
\label{sec:future_work}
Our $BFDR(q;a)$ considers tests which are independent across the $r_{j_t}$ for $j\in \{1,...,m\}$. \cite{Sun2009LargeScale} previously considered threshold selection in environs where the test statistics are dependent; using a notion termed the False Discovery Proportion (FDP). The results outlined in Section \ref{subsec:illustration} and the associated Theorem \ref{theorem2} could allow us to directly model correlations and limit their effect on performance. Additionally, further investigation of the $L_p$ regression model of \cite{LP_Norm} in the context of Theorem \ref{theorem2} is necessary. Optimal choices of $a$ based on adapting to the kurtosis scheme and having $a$ adapt to the set of tests $\{r_{1_t},...,r_{m_t}\}$ at each time point could admit improved performance globally across the full time series. 

Temporal dependence is also something that was not considered. Gaussian processes (\cite{Rasmussen2006Gaussian}) embedded in the direct modeling component of the posterior predictive could permit effective temporal modeling, and then in concordance with the thresholding results presented here could provide improved temporal performance.

\section{Conclusion}
\label{sec:conclusion}
Our work illustrates new connections between aleatoric uncertainty and decision theory, as well as new directions to be explored in working with thresholds in concordance with data-informed loss functions. In Section \ref{sec:decision theory} we construct a new loss designed for working with aleatoric uncertainty in the context of posterior predictive distributions. The theoretical results outlined in Section \ref{sec:data_informed}, and particularly section \ref{subsec:note_about_BFDR}, provide appealing theoretical justification for comparing certain classes of loss functions and multiplicity controlling threshold selection rules. Additionally, a new generalization of $BFDR(q)$ was introduced, with a proven connection to $L_p$ regression problems illustrated in Section \ref{subsec:illustration}, and a theorem was proven showing a way to vectorize computations to improve stability of performance of the computed BFDR in \ref{subsection:tensorized}. An outline of a manner in which to generalize the aleatoric detection model to general Gibbs-type nonparametric priors was discussed in Section \ref{sec:nonparametric}.

Simulations were done in a rigorous environment in Section \ref{sec:simulations}; across a range of value of $q$ and compared to the common environment of training on an individual time block and extrapolating to the broad range time points in the series. Performance is shown to be better nearly-uniformly in situations where the threshold is selected using the $BFDR(q;a=2)$ at each time step to modify thresholds temporally for multiplicity. The modified $BFDR(q;a)$ showed improvements in performance for small values of $q$ across the time series in the domain of Precision and Recall. Additionally the model performed adequately in balanced accuracy in comparison to strict threshold methods. However, the $BFDR(q;a)$ threshold suffered in accuracy, especially in high noise paradigms.

\section*{Acknowledgments}
We would like to thank the Microsoft Security Research Leadership Team for permission to publish.

\appendix
\section{Proof of \ref{lemma1}}

\begin{statement}
Suppose $f: \{\delta\}_{j=1}^m \in \{0,1\}^m \rightarrow q$ is a random function that is non-negative and piecewise-constant, monotonic in $\delta_j$ for some direction for each $\delta_j$; with induced randomness in $f$ coming from another random variable $\gamma$ which is a binary vector $\gamma_j \overset{ind.}{\sim} \sim\textrm{Bern}(r_j)$. Assume additionaly that $f$ is almost surely monotonic (in either direction) as a function of each $\gamma_j$. Suppose $g: q \rightarrow \eta = \argmax_\eta \{f(\eta): f(\eta) < q\}$.Then:

\begin{enumerate}
\item $\mathbb{E}\gamma[g(q;\gamma)] = \mathbb{E}\gamma[g; r_j](q) =  \argmax_\delta \{\mathbb{E}[f; r_j](q) : \mathbb{E}[f; r_j](q) < q\}]$
\item $\mathbb{E}\gamma[g; r_j]$ is interconnected with the monotonicity in $\gamma$ and $\delta$. If $f$ is monotone increasing (or decreasing) with respect to each $\gamma_j$, and monotone increasing (or decreasing) in the same direction with respect to each $\delta_j$, then the monotonicity of $\mathbb{E}\gamma[g; r_j]$ with respect to $\delta_j$ is the inverse of the monotonicity in $\delta_j$ before the expectation is taken. If they are in the opposite direction the monotonicity in $\delta_j$ is the same after the expectation is taken.  
\end{enumerate}
\end{statement}
\begin{proof}
\label{prooflemma1}

\begin{enumerate}

\item Note that $\mathbb{E}\gamma[g(q;\gamma)] = \mathbb{E}\gamma[\argmax_\delta \{f(\delta): f(\delta; \gamma) < q\}]$, and by the non-negativity of $\gamma_j$, and the non-negativity of $f$, we can exchange the expectation and $\argmax_\delta$; giving us:

\begin{equation}
\label{argmax}
\mathbb{E}\gamma[g(q;\gamma)] = \argmax_\delta \mathbb{E}\gamma[{f(q;\gamma): f(q; \gamma)< q}]
\end{equation}

Note that because $f$ is piecewise constant we get that $\mathbb{E}\gamma[f(\delta; \gamma)]$ is piecewise linear and a function of the parameters of each individual $\gamma_j$; which are the associated $r_j$. This gives us that $\mathbb{E}\gamma[f(\delta;\gamma] = \mathbb{E}[f; r_j](\delta)$.

This gives us that that our previous Equation \ref{argmax} can be represented as:

\begin{equation}
\mathbb{E}\gamma[g(q;\gamma)] = \argmax_\delta \{\mathbb{E}[f; r_j](q) : \mathbb{E}[f; r_j](q) < q\}]
\end{equation}
\item Examine a particular $\gamma\in\{\gamma_j\}_{j=1}^m$. Note that $\mathbb{E}_\gamma[f;r]$ is still a function taking $\{\delta_j\}_{j=1}^m \rightarrow q$, and that as $f$ is monotone (increasing or decreasing) in $\gamma$ this expected function is monotone (increasing or decreasing) in the same direction in the correspond $r$ mapping to $\gamma$. Note that as $r\rightarrow 1$ the function is increasing or decreasing in the direction of $\gamma$'s monotonicity, thus more likely to be greater than or less than $q$, if $f$ is monotone in the opposite direction with respect to $\delta$ corresponding to $\{\gamma,r\}$, then it preserves the monotonicity in $\delta$ to stay below threshold of $\mathbb{E}_\gamma[f; r](\delta) < q$. If it is monotone in same direction, the monotonicity in $\delta$ is inverted (changing signs allows the same arguement to apply for the opposite direction as well). 
\end{enumerate}
$\blacksquare$
\end{proof}

\section{Proof of \ref{theorem3}}
\label{prooftheorem3}
\begin{statement}
For:

\begin{equation}
\eta =  \argmax_{\eta\in \vec{K}} \{\sum_{j=1}^m \textrm{sign}(r_j-q) |(r_j - q)|^aI_{(r_j \leq \eta)} : \sum_{j=1}^m \textrm{sign}(r_j-q) |(r_j - q)|^aI_{(r_j \leq \eta)} < 0 \}
\end{equation}

where $\vec{K} = [0, \frac{1}{k}, \frac{2}{k},....,1]$ is some grid with resolution $k\in \mathbb{N}$, is difficult to compute using a loop. This can be greatly improved using vectorization.

Let $\vec{r} = \begin{pmatrix}
r_1 \\
r_2 \\
\vdots \\
r_m\end{pmatrix}$ and let $R= \vec{r}\otimes \vec{1}_{(k+1)}^T$. We construct an indicator matrix $\Psi$ corresponding to the following:

\begin{equation}
\Psi = [I(R_{jl} <(\vec{K}\otimes \vec{1}_{m}^T)^T_{jl})]_{jl}
\end{equation}

This matrix is of dimension $m\times (k+1)$. The $j$ index corresponds to the anomaly scores and the $l$ index corresponds to the grid. This can also be computing quickly through vectorization.

Then we have that the $BFDR(q;a)$ can be computed as follows:

\begin{equation}
\eta = \frac{\textrm{\textit{max index}}_l \{\vec{1}_m^T[\textrm{sign}(R\odot \Psi - q\times\Psi)\odot |R\odot \Psi - q\times\Psi|^a] < 0 \}}{k}
\end{equation}

Where the sign function, absolute value, and exponentiation are taken elementwise.

\end{statement}

\begin{proof}

Note that $\Psi_{jl} = I_{(r_j < K_l)}$ and that $R_{jl}$ for $R= \vec{r}\otimes \vec{1}_{(k+1)}^T$ is just $r_j$ no matter the value of $l$. 

Thus:

\begin{equation}
[\textrm{sign}(R\odot \Psi - q\times\Psi)\odot |R\odot \Psi - q\times\Psi|^a]_{jl} = \textrm{sign}(r_j-q) |(r_j - q)|^aI_{(r_j \leq \vec{K}_l)}
\end{equation}

This is because the $\Psi$ term is just the indicator function getting mapped associatedly to each element of $R$, and the term $q\times\Psi$ making sure we are only subtracting the FDR control $q$ for corresponding entries in the matrix $R\otimes \Psi$ which are nonzero. The left multiplication by $\vec{1}_m^T$ equates to the sum of $\sum_{j=1}^m$  in the original equation. This gives us a column vector of dimension $k$ where the $l$ index corresponds to the $l$'th element of $K$. 

We're trying to find the maximal index such that this vector is less than zero, corresponding to the $\argmax_{\eta \in\vec{K}}$. The division by $k$ is a scaling to map the index $l$ to $\frac{l}{k} = K_l$ which is the solution to the maximization problem.

\end{proof}
\section{Additional Performance of Tensorized Versus Looped \texorpdfstring{$BFDR(q;a)$}{BFDR(q;a)} On Clock Times}

\label{appendix:clock_time}

\begin{figure}
        \centering
        \includegraphics[width=0.7\textwidth]{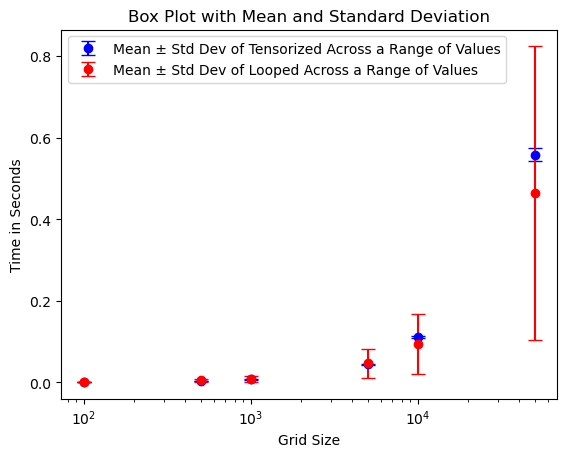}
        \caption{Boxplots of Clock times across 1500 replications for the setting outlined in this section. Scale is logarithmic showing that both the looped and vectorized versions are linear in $k$, however, the looped version has much higher standard deviation; increasing as a function of $k$. This is likely attributable to the discussion introduced in Section \ref{subsection:tensorized}.}
        \label{fig:clocktime_boxplot}
\end{figure}

We repeat the replications test introduced in Section \ref{subsection:tensorized}, across a range of different values of $k$, and assess performance. Again, $q=.2$, $a=2$, and $k$ is evaluated for the values $[100,500,1000,5000,10000,50000]$. Results are illustrated in Figure \ref{fig:clocktime_boxplot}.

%Bibliography
\bibliographystyle{unsrt}  
\bibliography{DTLSOD_BFDR}

\end{document}